\documentclass[aps,amsmath,amssymb]{revtex4}
\usepackage{graphicx}
\usepackage{wrapfig}
\usepackage{subfigure}
\usepackage{mathptmx}
\usepackage{psfig}
\usepackage{dcolumn}
\usepackage{bm}
\usepackage{tabls}
\newcommand{\be}{\begin{eqnarray}}
\newcommand{\ee}{\end{eqnarray}}
\begin{document}
\title{Kinks in Discrete Light Cone Quantization}
\author{Dipankar Chakrabarti}
\email{dipankar@theory.saha.ernet.in}
\author{A. Harindranath}
\email{hari@theory.saha.ernet.in}
\affiliation{Theory Group, Saha Institute of Nuclear Physics \\
 1/AF Bidhan Nagar, Kolkata 700064, India}
\author{ L$\!\!$'ubomir Martinovi\v c}
\email{fyziluma@savba.sk}
\affiliation{Institute of Physics, Slovak Academy of Sciences \\
D\'ubravsk\'a cesta 9, 845 11 Bratislava, Slovakia}
\author{J. P. Vary}
\email{jvary@iastate.edu}
\affiliation{Department of Physics and Astronomy, Iowa State University,
Ames, IA 50011, U.S.A.}
\date {September 30, 2003}
\begin{abstract}
We investigate  non-trivial topological structures in Discrete Light Cone 
Quantization (DLCQ) through the example of the broken symmetry phase of 
the two dimensional $\phi^4$ theory using anti periodic boundary condition (APBC). We present evidence for 
degenerate ground states which is both a  
signature  of spontaneous symmetry breaking and mandatory for the 
existence of kinks. Guided by a constrained variational calculation 
with a coherent state ansatz, we then extract the vacuum  energy 
and kink mass and compare with classical and semi - classical results. 
We compare the DLCQ results for the number density of bosons in the kink 
state and the Fourier transform of the form factor of the kink  with
corresponding observables in the  coherent variational kink state.

\end{abstract}
\maketitle
\noindent {\bf 1. Introduction}

Motivated by the remarkable work of  Rozowsky and Thorn \cite{RT}, 
we have recently investigated \cite{pp1}  the broken symmetry phase
of two dimensional $\phi^4$ theory in DLCQ \cite{dlcq} with periodic 
boundary condition
(PBC) without the zero momentum mode. Using a coherent 
state variational calculation
as a guide, we extracted the vacuum energy density 
and kink mass from the
results of matrix diagonalization. We also presented the Fourier transform
of the form factor of the lowest excitation as well as the number density of
elementary constituents of that state. Since the zero momentum mode was dropped
in these investigations \cite{RT,pp1},
the lowest state appeared as a kink-antikink pair because of the periodic
boundary condition which
   implies that we are working in the sector with topological charge
   equal to zero. The results from these studies are not free from 
ambiguity at least in the finite volume
because of the potential role played by the constrained zero momentum mode. 
With anti periodic boundary condition (APBC), the zero 
momentum mode is absent and hence
calculations are free from the ambiguity created when it is simply 
neglected. With APBC one expects the ground state to be a kink or an antikink.

The quantum kink on the light front was addressed first by 
Baacke \cite{Baacke}
in the context of semi-classical quantization. As Baacke indicated, light
front quantization offers the advantage of preserving translational
invariance. To extract the kink mass in the classical theory he 
approximately diagonalized the
mass operator ($M^2 = P^+P^-$). 
He pointed out the advantage of light front
quantization in handling the translation mode.    

In this work we address the problem of the Fock space description of the 
topological structure in quantum field theory 
in the context of kinks that appear in the broken symmetry phase of two 
dimensional $\phi^4$ theory. As further background, it is worthwhile to 
recall
that the study of these objects in lattice field theory is also highly
non-trivial \cite{Ciria,Ardekani}. 
Within our own approach, we must qualify results by uncertainty due to
unknown artifacts arising from discretization.

\noindent {\bf 2. Notation and Conventions}

We start from the Lagrangian density 
\be
{\cal L} = \frac{1}{2} \partial^\mu \phi \partial_\mu \phi +
 \frac{1}{2} \mu^2 \phi^2 -  \frac{\lambda}{4!} \phi^4.
\ee
The light front variables are defined by $ x^\pm = x^0 \pm x^1$. 

The Hamiltonian density 
\be
{\cal P}^- =
 - \frac{1}{2} \mu^2 \phi^2 +  \frac{\lambda}{4!} \phi^4
\ee
defines the Hamiltonian 
\be
P^- & = & \int dx^- {\cal P}^-~
     \equiv  \frac{L}{2 \pi} H
\ee
where $L$ defines our compact domain $ - L \le x^- \le +L$. Throughout this
work we address the energy spectrum of $H$. 

The longitudinal momentum operator is 
\be
P^+ & = & \frac{1}{2} \int_{-L}^{+L} dx^- \partial^+ \phi \partial^+ \phi
 \equiv  \frac{2 \pi}{L}K
\ee
where $K$ is the dimensionless longitudinal momentum operator. The mass
squared operator $M^2 = P^+ P^- = KH$.

In DLCQ with APBC, the field expansion has the form 
\be
\Phi(x^-) = \frac{1}{\sqrt{4 \pi}} \sum_n \frac{1}{\sqrt{n}} 
\left [a_n e^{-i \frac{n \pi}{L} x^-} + a_n^\dagger e^{i \frac{n
\pi}{L} x^-} \right ]. 
\ee  
Here $ n = \frac{1}{2}, \frac{3}{2},\ldots. $


The normal ordered Hamiltonian  is given by

\be
H & = & - \mu^2 \sum_n \frac{1}{n} a_n^\dagger a_n
+ \frac{\lambda}{4 \pi} \sum_{k \le l, m\le n}~ \frac{1}{N_{kl}^2}
\frac{1}{\sqrt{klmn}}
 a_k^\dagger a_l^\dagger a_n a_m
\delta_{k+l, m+n} \nonumber \\
&~& + \frac{\lambda}{4 \pi} \sum_{k, l \le m\le n}~ \frac{1}{N_{lmn}^2}
  \left [
a_k^\dagger a_l a_m a_n + a_n^\dagger a_m^\dagger a_l^\dagger a_k \right ]~
\delta_{k, l+m+n}
\ee
with
\be N_{lmn} & = & 1 ,~ l \ne m \ne n, \nonumber \\
            & = & \sqrt{2!},~ l=m \ne n, ~ l \ne m=n,\nonumber \\
            & = & \sqrt{3!},~ l=m=n,
\ee
and
\be
N_{kl} & = & 1, k \ne l,~ \nonumber \\
       & = & \sqrt{2!},~ k=l.
\ee

\noindent {\bf 3. Coherent State Calculations}

Rozowsky and Thorn \cite{RT} carried out a coherent state 
variational calculation for DLCQ
in the case of PBC without the zero 
momentum mode. 
In this section we carry out the analogous calculation for APBC. 
The result of this calculation, being semi-classical, is especially reliable in the
weak coupling region and we can use its functional form  to extract 
the kink  mass
from the numerical results of matrix diagonalization.    

Choose as a trial state, the coherent state
\be
\mid \alpha \rangle = {\cal N} e^{\sum_{n} \alpha_n a_n^\dagger} \mid 0
\rangle
\ee
where ${\cal N}$ is a normalization factor. 

With APBC we have
\be
\frac{\langle \alpha \mid \phi (x^-) \mid \alpha \rangle}{\langle \alpha
\mid \alpha \rangle}  =  \frac{1}{\sqrt{4 \pi}} ~ f(x^-) 
\ee
with 
\be
f(x^-) = \sum_{m=1}^N
\frac{1}{\sqrt{m - \frac{1}{2}}} \Big [ \alpha_{m- \frac{1}{2}} 
e^{-i \frac{\pi}{L}(m -\frac{1}{2}) x^-}+ \alpha_{m- \frac{1}{2}}^{*}
e^{i \frac{\pi}{L}(m -\frac{1}{2}) x^-} 
\Big ]~.  \label{uvcf}
\ee
Minimizing the expectation value of the Hamiltonian, we obtain
\be
f_{min} = \pm \sqrt{\frac{24 \pi \mu^2}{\lambda}} = \pm \sqrt{\frac{3}{g}}.
\ee
Set 
\be
f(x^-) & = & \sqrt{\frac{3}{g}},~~ 0 < x^- < L~, \nonumber \\
          & = & - \sqrt{\frac{3}{g}},~~ - L  < x^- < 0~.
\ee
Then we get 
\be
\alpha_{m- \frac{1}{2}} = \sqrt{\frac{3 }{g}}~ \frac{i}{\pi} ~
\frac{1}{\sqrt{m - \frac{1}{2}}}~, ~~ m=1, 2, 3, \ldots ,
\ee
and
\be
f(x^-) = \frac{2}{\pi} \sqrt{\frac{3}{g}} \sum_{j}
\frac{1}{j}~ \sin~\frac{j\pi x^-}{L}~ \label{uvf}
\ee
where $ j = \frac{1}{2}, \frac{3}{2}, \frac{5}{2}, $ etc.
The number density of bosons with momentum fraction $x (= \frac{j}{K})$
is given by
\be
\chi(x) &=& \frac{\langle \alpha \mid a_j^\dagger a_j \mid \alpha \rangle} 
{\langle \alpha \mid \alpha \rangle}
 =  \alpha_j^2
\ee
where $ \alpha_j 
\sim \frac{1}{\sqrt{j}}$.

In this case we get

\be
\frac{1}{\langle \alpha \mid \alpha \rangle}
\frac{2 \pi}{L} \int dx^- \langle \alpha \mid \phi^2(x^-) \mid \alpha \rangle 
 =  \frac{8}{\pi^2} \frac{3}{g} \sum_j \frac{1}{j^2}
\ee
where $ j = 1, 3, 5, \ldots .$
In the unconstrained variational calculation for PBC, the expectation 
value of the
longitudinal momentum operator is infinite since $ f(x^-)$ is discontinous
at $x^-=0$.  To cure this deficiency,
Rozowsky and Thorn performed a constrained variational calculation. 
Here we provide an outline of the analogous calculation for APBC.  
 For constrained variational calculation in the case of PBC with the 
inclusion of a zero mode, see Ref. \cite{sugi}.  

With $ \langle K \rangle = \frac{L}{2 \pi} \frac{\langle \alpha \mid P^+ \mid \alpha \rangle}
{\langle \alpha \mid \alpha \rangle} $, and
$ f' = \frac{\partial f(x^-)}{\partial x^-}$ we have
\be
K = \frac{L}{4 \pi^2} \int_{-L}^{+L} dx^- (f')^2~. \label{me}
\ee
Minimizing
\be
 \frac{1}{\mu^2}~\frac{\langle \alpha \mid H_\beta \mid \alpha \rangle}
{\langle \alpha \mid
\alpha \rangle} 
= \frac{1}{ L} \int_{-L}^{+L} dx^-
\Big [  \beta \Big \{ \frac{L^2}{4 \pi^2} (f')^2 - \langle K \rangle L \Big \}
- \frac{1}{4}  f^2 + \frac{\lambda}{192 \mu^2} f^4 \Big ]
\ee
 we obtain
\be
- 2 \beta \frac{L^2}{4 \pi^2} \frac{\partial^2 f}{\partial (x^-)^2}
- \frac{1}{2} f + \frac{\lambda}{48 \pi \mu^2} f^3 =0.
\ee
Putting $ f(x^-) = f_0 F(u) $ where the variable
 $ u =\frac{ 2 x^- + L}{L}{\overline K} $ with
\be {\overline K} = {\overline K}(k) = \int_0^1 dt
(1-t^2)^{-\frac{1}{2}} (1-k^2 t^2)^{-\frac{1}{2}},
\ee
 we have,
\be
\frac{\partial^2 F}{\partial u^2} = - \frac{1}{4 {\overline K}^2 \beta} F +
\frac{\lambda f_0^2}{ 96 {\overline K}^2 \beta \pi \mu^2} F^3.
\ee
Comparing with the differential equation satisfied by the Jacobi Elliptic
Function  $ sn(u,k)$, namely,
\be
\frac{\partial^2 sn(u,k)}{\partial u^2} = -(1+k^2) ~ sn(u,k)
+ 2 k^2 ~ sn^3(u,k),
\ee
we get
\be
f(x^-) = f_0~ sn \left(\frac{x^-}{L}{\overline K}, k \right) 
\ee
with
\be
\beta = \frac{1}{4 {\overline K}^2 (1+k^2)} ~~ {\rm and}~~ f_0^2 =
\frac{48 k^2 \pi \mu^2}{ \lambda (1+k^2)}.
\ee
Note that we have imposed APBC on the solution.
By explicit calculation  we get
\be
\langle K \rangle =  \frac{8 \mu^2}{ \pi \lambda} {\overline K} \Big [ E(k) - \frac{1-k^2}{1+k^2}
{\overline K}(k) \Big ]
\ee
with 
\be
E(k) = \int_0^1 dt ~ \frac{\sqrt{1 - k^2t^2}}{\sqrt{1-t^2}}
\ee

and
\be
\frac{\langle \alpha \mid H \mid \alpha \rangle}
{\langle \alpha \mid \alpha \rangle} = - \frac{24 k^2 \pi \mu^4}{\lambda
(1+k^2)^2}
+ \frac{64 \mu^6}{\lambda^2 (1+k^2)\langle K \rangle }
\Big [ E(k) - \frac{1-k^2}{1+k^2} {\overline K}(k)
\Big ]^2.
\ee
In the $ \langle K \rangle \rightarrow \infty$ limit, $ k \rightarrow 1 $ and we get
\be
\frac{\langle \alpha \mid H \mid \alpha \rangle}
{\langle \alpha \mid \alpha \rangle}= - \frac{6 \pi \mu^4}{\lambda} +
\frac{32 \mu^6}{\lambda^2 \langle K \rangle}~.  \label{gev}
\ee
Interpreting the state $ \mid \alpha \rangle$ to be a kink state, 
we identify the first term as the vacuum energy
density which is the classical vacuum energy density.
The second term is identified as $ \frac { M_{kink}^2}{\langle K \rangle }$.
Then we get the classical kink mass $ M_{kink
} =
\frac{4 \sqrt{2} \mu^3}{\lambda} $.

Using the Fourier expansion \cite{AS}
\be
sn(u,k) = \frac{1}{{\overline K}}\frac{2 \pi}{\sqrt{k^2}}~ \sum_{m=1}^\infty ~\frac{q^{m - {1 \over
2}}}{1 - q^{2m -1}} ~ \sin \frac{(2m-1) \pi u}{2{\overline K}}
\ee
where $ q = exp \left({-\pi \frac{{\overline K}(1-k^2)}
{{\overline K}(k^2)}}\right) $
we have 
\be
f(x^-) = \frac{2 \pi}{{\overline K}}~ \sqrt{\frac{48 \pi
\mu^2}{\lambda(1+k^2)}}~
\sum_{j} ~\frac{q^{j}}{1 - q^{2j}} ~ \sin \frac{j \pi x^-}{L}~. \label{cvf}
\ee
In the limit $ k^2 \rightarrow 1$, using $ q \rightarrow limit_{k^2 \rightarrow 1}
\left (1 - \pi ~\frac{{\overline K}(k^2-1)}{{\overline K}(k^2)}\right ) $ 
so that 
$ (1 - q^{2m-1}){\overline K} \rightarrow (2m-1) \frac{\pi^2}{2}$ since
${\overline K}(0)=\frac{\pi}{2}$, it is readily verified that in the limit $k^2
\rightarrow 1$, the expression for $f(x^-)$ in the constrained variational
calculation given by Eq. (\ref{cvf}) goes over 
to that in the
unconstrained variational calculation given by Eq. (\ref{uvf}). 

\noindent {\bf 4. Fourier transform of the form factor in DLCQ} 

An observable that yields considerable insight for the spatial
structure of the topological object is the Fourier transform of its form
factor. We compute the Fourier transform of the form
factor of the lowest state which, according to Goldstone and Jackiw
\cite{gj},
in the weak coupling (static) limit,
represents the kink profile. Let $ \mid K \rangle$ 
and $ \mid K' \rangle$ denote this state
with momenta $K$ and $K'$.
In the continuum theory,
\begin{eqnarray}
\int_{- \infty}^{+ \infty} dq^+ exp\{- \frac{i}{2} q^+ a\}
\langle K' \mid \Phi(x^-) \mid K \rangle = \phi_c(x^- - a).
\end{eqnarray}
In DLCQ, we diagonalize the Hamiltonian for a given $K = \frac{L}{2 \pi}P^+$. 
For the computation
of the form factor, we need the same state at different $K$ values since $
K'=K+q$. We
proceed as follows. We diagonalize the Hamiltonian, 
say, at $K=41$ (even particle sector). We diagonalize the
Hamiltonian at the neighboring  $K$ values, 
$K=40.5, 41.5, 39.5, 42.5, 38.5,
43.5, 37.5, 44.5, 36.5, 45.5$ (odd particle sectors). 
In this particular example, the dimensionless momentum 
transfer
ranges from $-4.5$ to $+4.5$. 
 If $K$ is large enough to be near the continuum, then, in the
spontaneous symmetry broken phase, with degenerate even and odd states, we
can be confident that all these lowest states correspond to the {\em same}
physical state observed at different longitudinal momenta. The test that the states are degenerate is that they have the
same $M^2$, so the eigenvalues of $H$ fall on a linear trajectory as a
function of $\frac{1}{K}$.  

We proceed to compute the matrix element of the field
operator between the lowest state at $K=40$ and the other specified values 
of $K$ and sum the amplitudes which corresponds to the choice of the shift parameter $a=0$. 
In summing the amplitudes, we need to be careful about the phases. 
First we note that $K$ is a conserved quantity, so eigenfunctions at
different $K$ values have an independent arbitrary complex phase factor.
To fix the phases, we accept the guidance of the coherent state analysis.
We set the overall sign of the lowest states for all $K$ values such 
that the matrix elements $\langle K+n \mid a_n^\dagger \mid K \rangle~ = ~
positive$ and $\langle K-n \mid a_n \mid K \rangle ~=~ negative$.  
In addition, there is one 
overall complex phase that we apply to the profile function so that it is real at the boundaries.  
That the sum of all
terms for the profile function produces the shape of a kink, with very small
imaginary component, is nevertheless a non-trivial result.  
It is a further non-trivial result that the magnitude of the kink represents 
a physically sensible result.

\noindent {\bf 5. Numerical Results}

With APBC,  for integer (half integer)  values of $K$ we
have even (odd) number of particles. 
The dimensionality of the matrix
in the even and odd sectors for different values of $K$ is presented in Table
\ref{dim}.
All results presented here were obtained on small
clusters of computers ($\le$ 15 processors) using  
the Many Fermion Dynamics (MFD) code adapted to bosons \cite{mfd}. The Lanczos diagonalization
method is used in a highly scalable algorithm allowing us to proceed to
sufficiently high values of K to numerically observe the phenomena we
sought.

\begin{figure}[hbtp]
\includegraphics[width=6in,clip]{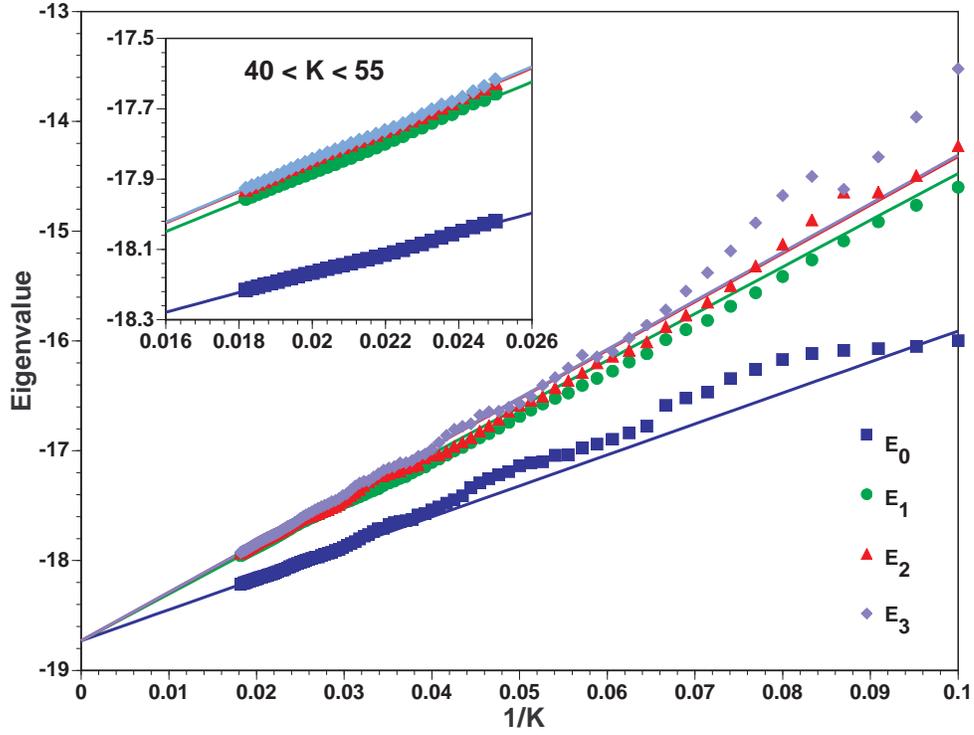}
    \caption{Lowest four eigenvalues for even and odd sectors 
as a function
of $\frac{1}{K}$ for $\lambda$=1.0. The inset shows the details over the
range $ 40 \le K \le 55$. The  discrete  points
are the DLCQ eigenvalues while the straight lines are the linear fits to
the $ 40 \le K \le 55$ data constrained to have the same intercept.}
\label{4evs1}

\end{figure}

Since the Hamiltonian exhibits the $ \phi \rightarrow - \phi$ symmetry, the
even and odd particle sectors of the theory are decoupled.  With a positive
$\mu^2$, at weak coupling, the lowest state in the odd particle sector is a
single particle carrying all the momentum. In the even particle sector, the lowest
state consists of two particles. Thus for massive
particles, there is a distinct mass gap between odd and even particle
sectors. With a negative $\mu^2$, at weak coupling, the situation is drastically different.
Now, the lowest states in the odd and even particle sectors consist of
the maximum number of particles carrying the lowest allowed momentum. Thus, in
the continuum limit, the possibility arises that the states in the even and odd
particle sectors become degenerate. 
A clear signal of SSB is the degeneracy of the
spectrum in the even and odd particle sectors. 
Thus at finite $K$,  
we can compare the spectra for an integer K value (even particle sector) 
and its neighboring half
integer $K$ value (odd particle sector) and look for 
degenerate states. In Fig. \ref{4evs1} we show the lowest four energy 
eigenvalues  in the broken symmetry
phase for the even and odd particle sectors for $\lambda$=1.0    
as a function 
of $\frac{1}{K}$. The points represent results at half integer increments in
$K$ from $K=10$ to $K=55$. The overall trend is towards smoother
behavior at higher $K$. There is an apparent small oscillation superimposed
on a generally linear trend for each state. We believe that the
oscillations represent an artifact of discretization. These 
oscillations decrease with
increasing $K$. The smooth curves in Fig. \ref{4evs1} are linear fits to
the eigenvalues in the range from $K=40$ to $K=55$ constrained to have 
the same intercept.  

With guidance from the constrained variational calculation, see
Eq. (\ref{gev}),  we can extract the kink mass from the linear fit to 
the DLCQ data for
the ground state eigenvalue. We fit the
$ \lambda=1.0$ data in the range $ 40 \le K \le 55$
to a linear form ($ C_1+C_2/K$). 
There are two reasons for this choice: (1) this is the maximum
amount of data for which the $K$-artifacts seem reasonably absent; (2)
independent fits and extrapolations from the four lowest eigenvalues are
very close to each other at $ K \rightarrow  \infty$.
We quote $C_1$ as the vacuum energy density and $C_2$ as the kink 
mass in Table \ref{comp}. 
We obtain the uncertainties from the spread in these results 
arising from constrained fits to subsets of the data in this same range.
For comparison, the corresponding classical values 
(classical vacuum energy density ${\cal E}=- 6 \pi       
\mu^4/\lambda$)
are also
presented. The agreement appears reasonable.

\begin{figure}[hbtp]
\centering
\includegraphics[width=6in,clip]{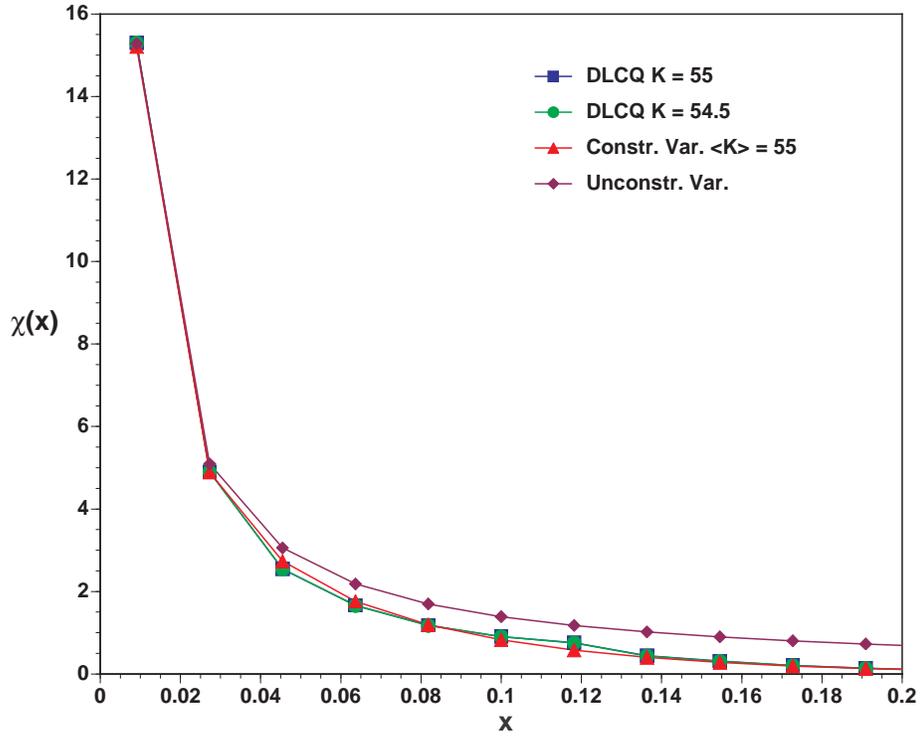}
    \caption{The number density $\chi(x)$  for even ($K=55.0$)
and odd ($K=54.5$)
sectors for  $\lambda=1$ compared with unconstrained 
and constrained 
($\langle K \rangle=55$) variational results.}
\label{nxs}
\end{figure}

Next we examine the behavior of the number density $\chi(x)$ for
the kink state. In the broken phase, the ground states in the even
and odd particle sectors are degenerate in the continuum limit. 
In Fig. \ref{nxs} we show $\chi(x)$ for
$K=55$ and $K=54.5$ for $\lambda$ = 1. For this coupling
the number densities for even and odd sectors are almost identical
to each other indicative of degenerate states. In Fig. \ref{nxs}, we 
also compare the DLCQ number density  with  
that predicted by the unconstrained and constrained 
variational
calculations. At sufficiently large $K$ and low $\lambda$, they appear
to agree at a level which is reasonable for the comparison of a quantal
result with a semi-classical result.

\begin{figure}[hbtp]
\centering
\includegraphics[width=6in,clip]{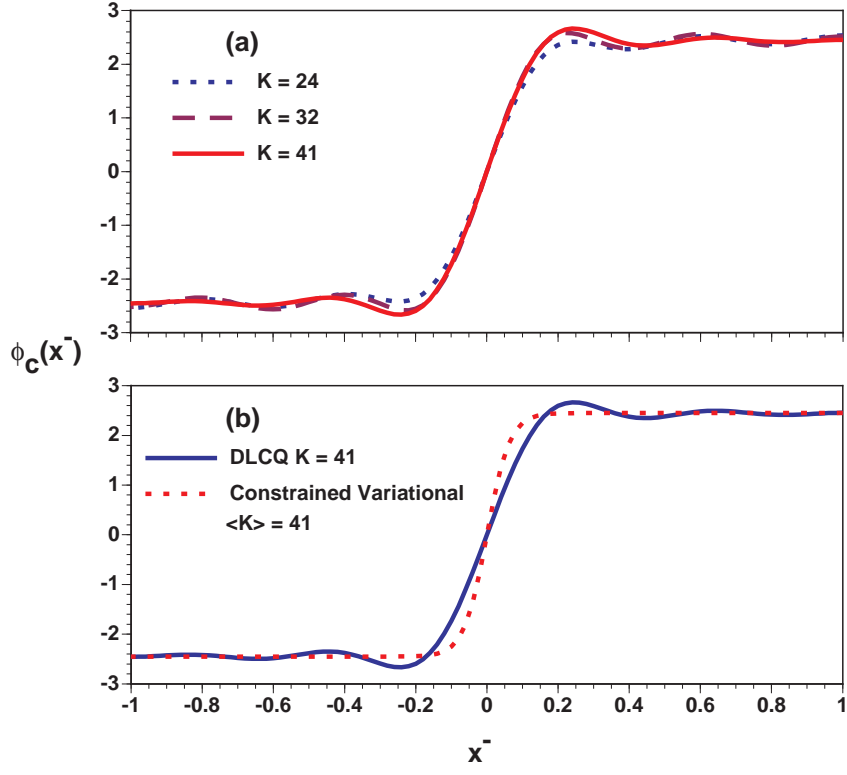}
\caption{Fourier Transform of the kink form factor at
$\lambda$=1; (a) results for  $K=24, 32$, and $41$ each obtained with DLCQ
eigenstates from 11 values of $K$ centered on the designated $K$ value; (b)
comparison of DLCQ profile at K=41 with constrained variational result with
$ \langle K \rangle=41$. 
}
\label{profile}
\end{figure}

Following Goldstone and Jackiw, we have calculated the Fourier transform of
the form factor of the kink state in DLCQ at weak coupling. 
In Fig. \ref{profile}(a) we show the profile 
calculated in DLCQ for $\lambda=1$ at three selected $K$ values. It is
clear that at $\lambda=1$ the profile is that of a kink which appears
reasonably converged with increasing $K$. In Fig. \ref{profile}(b) 
we compare the $K=41$ DLCQ profile with that of a constrained 
variational coherent
state calculation of Eq. (\ref{cvf}) with $ \langle 
K \rangle =41$. In the
unconstrained variational calculation, this function is discontinuous
at $x^-=0$ and $ \langle K \rangle $, the expectation value of the dimensionless longitudinal 
momentum operator, is infinite. In the variational calculation where
$ \langle K \rangle $ is constrained to be finite, the kink profile is a 
smooth function of
$x^-$ as seen in Fig. \ref{profile}(b).  In the limit $ \langle 
K \rangle \rightarrow \infty$, the kink profile from
constrained variational calculation approaches that of the
unconstrained case. For each $K$ shown, we utilize 11 sets of DLCQ results
to construct the profile function. Thus, for $K=41$ we employ results at
$K=41$ and at $K=36.5$ through 45.5 in unit steps.

To summarize,  we have 
demonstrated the existence of degenerate lowest eigenstates in two
dimensional $\phi^4$ theory in DLCQ with APBC. The degeneracy of energy levels  
is both a signature of spontaneous symmetry breaking and essential for the
existence of kinks.  Using the 
constrained variational calculation as a guide, we
have extracted the vacuum  energy density and the kink mass 
for $\lambda=1$.  
We have extracted the number density of bosons in the kink state 
and compared it with predictions from coherent state variational
calculations. We have also calculated the Fourier transform of the form
factor of the kink and compared it with its counterpart in the variational
approach. We interpret these results as indicative of the viability of DLCQ
for addressing non-trivial phenomena in quantum field theory.

This work is supported in part by the Indo-US
Collaboration project jointly funded by the U.S. National Science
Foundation (INT0137066),  the Department of Science and Technology, India
(DST/INT/US (NSF-RP075)/2001). This work is also supported in part by the US
Department of Energy, Grant No. DE-FG02-87ER40371, Division of
Nuclear Physics and one of the  authors (L.M.) was partially 
supported by the VEGA grant No.2/3106/2003.
Two of the authors (D.C and A.H)
acknowledge helpful discussions with Asit De and Samir Mallik.

\eject
\tablinesep=.1in
\arraylinesep=.1in
\extrarulesep=.1in
\begin{table}
\begin{tabular}{||c|c||c|c||}
\hline \hline
\multicolumn{2}{||c||}{odd sector} & \multicolumn{2}{c||}{even sector} \\
\hline 
  K  & dimension & K & dimension \\
\hline
15.5 &  295  & 16 & 336    \\
 31.5 & 12839 & 32 & 14219  \\
39.5 & 61316 & 40 & 67243    \\
44.5 &  151518 & 45 & 165498  \\ 
49.5 & 358000  & 50 & 389253  \\
54.5 & 813177 & 55 & 880962   \\
\hline\hline
\end{tabular}
\caption{Dimensionality of the Hamiltonian matrix in odd and even particle
sectors with anti periodic boundary condition.}
\label{dim}
\end{table}

\tablinesep=.1in
\arraylinesep=.1in
\extrarulesep=.1in
\begin{table}[bht]
\vskip 1cm
\centering
\begin{tabular}{||c|c|c|c|c|c||}
\hline \hline
   $\lambda$   & \multicolumn{2}{c|}{vacuum energy} &
\multicolumn{3}{c||}{soliton mass}\\
\hline
 & classical & DLCQ & classical & semi-classical & DLCQ\\
\hline
1.0 & -18.85  & -18.73 $ \pm 0.05$ &  5.66  & 5.19 & 5.3 $ \pm 0.2$  \\
\hline
\hline
\end{tabular}
\caption{Comparison of vacuum energy density and soliton mass extracted from
the continuum limit  of our DLCQ data, with classical results. For soliton
mass, the semi-classical result \cite{dashen} is also shown. }
\label{comp}
\end{table}


\eject

\end{document}